\documentclass[aps,prl,reprint,amsmath,amssymb,superscriptaddress]{revtex4-1}

\usepackage[utf8]{inputenc}
\usepackage{textcomp}
\usepackage{graphicx}
\graphicspath{{img/}}
\usepackage{hyperref}
\usepackage[all]{hypcap}
\hypersetup{
  colorlinks,%
  citecolor=blue,%
  filecolor=blue,%
  linkcolor=blue,%
  urlcolor=blue
}

\usepackage{soul}
\soulregister\cite{7}
\soulregister\eqref{7}

\newcommand*{\bra}[1]{\ensuremath{\langle #1 \vert}}
\newcommand*{\ket}[1]{\ensuremath{\vert #1 \rangle}}
\newcommand*{\inner}[2]{\langle #1 | #2 \rangle}

\renewcommand*{\eqref}[1]{Eq.~(\ref{#1})}

\newcommand*{\figref}[1]{Fig.~\ref{#1}}
\newcommand*{\Figref}[1]{Figure~\ref{#1}}

\begin{document}

\title{Quantum Sampling Algorithms for Near-Term Devices}
\date{\today}

\author{Dominik S.~Wild}
\affiliation{Max Planck Institute of Quantum Optics, Hans-Kopfermann-Stra{\ss}e 1, D-85748 Garching, Germany}

\author{Dries Sels}
\affiliation{Center for Computational Quantum Physics, Flatiron Institute, New York, New York 10010, USA}
\affiliation{Department of Physics, New York University, New York, New York 10003, USA}

\author{Hannes Pichler}
\affiliation{Institute for Theoretical Physics, University of Innsbruck, Innsbruck A-6020, Austria}
\affiliation{Institute for Quantum Optics and Quantum Information, Austrian Academy of Sciences, Innsbruck A-6020, Austria}

\author{Cristian Zanoci}
\affiliation{Department of Physics, Massachusetts Institute of Technology, Cambridge, Massachusetts 02139, USA}

\author{Mikhail D.~Lukin}
\affiliation{Department of Physics, Harvard University, Cambridge, Massachusetts 02138, USA}

\begin{abstract}
  Efficient sampling from a classical Gibbs distribution is an important computational problem with applications ranging from statistical physics over Monte Carlo and optimization algorithms to machine learning. We introduce a family of quantum algorithms that provide unbiased samples by preparing a state encoding the entire Gibbs distribution. We show that this approach leads to a speedup over a classical Markov chain algorithm for several examples including the Ising model and sampling from weighted independent sets of two different graphs. Our approach connects computational complexity with phase transitions, providing a physical interpretation of quantum speedup. Moreover, it opens the door to exploring potentially useful sampling algorithms on near-term quantum devices as the algorithm for sampling from independent sets on certain graphs can be naturally implemented using Rydberg atom arrays.
\end{abstract}

\maketitle

Efficient algorithms that sample from Gibbs distributions are of broad practical importance in areas including statistical physics~\cite{Landau2013}, optimization~\cite{Kirkpatrick1983}, and machine learning~\cite{Bishop2006}. Quantum systems are naturally suited for encoding sampling problems: according to the Born rule, a projective measurement of a quantum state $\ket{\psi}$ in an orthonormal basis $\{\ket{s}\}$ yields a random sample drawn from the probability distribution $p(s) = | \inner{s}{\psi} |^2$. This observation underpins recent work aiming to demonstrate quantum advantage by sampling from a probability distribution defined in terms of a quantum gate sequence~\cite{Bouland2018} or an optical network~\cite{Aaronson2011}. While these efforts have led to impressive experimental demonstrations~\cite{Arute2019,Zhong2020}, thus far they have limited implications for practically relevant problems. In this Letter, we introduce a family of quantum algorithms for sampling from classical Gibbs distributions. We illustrate our approach with several specific examples including sampling from the Gibbs distribution of the Ising model and sampling from weighted independent sets. Since approximating the size of the maximum independent set on a random graph is NP hard~\cite{Garey1979}, the latter encompasses computationally hard problems relevant for practical applications~\cite{Bomze1999,Butenko2006,Fortunato2010}. In contrast to many of the pioneering quantum algorithms for sampling problems~\cite{Lidar1997,Terhal2000,Somma2007,Somma2008,Wocjan2008,Bilgin2010,Yung2010,Temme2011,Riera2012,Yung2012,Boixo2015,Ge2016,Swingle2016,Harrow2020}, our approach does not require a large-scale, universal quantum computer and may, in certain instances, be realized on near-term quantum devices using highly excited Rydberg states~\cite{PRA}.

\begin{figure}[b]
  \centering
  \includegraphics{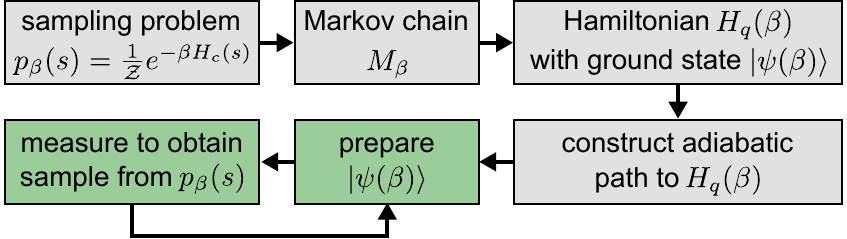}
  \caption{Key steps in the construction of the proposed quantum algorithms. The green boxes constitute the sampling procedure, which is carried out on a quantum computer.}
  \label{fig:fig1}
\end{figure}

The key ideas of our work are summarized in \figref{fig:fig1}. We focus on classical spin models, labeling a spin configuration by $s = s_1 s_2 \dots s_n$ with each spin being either up ($s_i = + 1$) or down ($s_i = -1$). The desired Gibbs distribution $p(s) = e^{- \beta H_c(s)}/\mathcal{Z}$ is defined in terms of the energies $H_c(s)$ and the inverse temperature $\beta$ with $\mathcal{Z} = \sum_s e^{- \beta H_c(s)}$ denoting the partition function. Sampling from a classical Gibbs distribution can be reduced to preparing the quantum state
\begin{equation}
  \ket{\psi(\beta)} = \frac{1}{\sqrt{\mathcal{Z}}} \sum_s e^{- \beta H_c(s)/2} \ket{s},
  \label{eq:gibbs_state}
\end{equation}
which we refer to as the Gibbs state, followed by a projective measurement in the $\{ \ket{s} \}$ basis. To prepare this state, we start from a classical Markov chain Monte Carlo algorithm for sampling from the Gibbs distribution. Any such Markov chain can be mapped onto a so-called parent Hamiltonian $H_q(\beta)$ with $\ket{\psi(\beta)}$ a ground state~\cite{Verstraete2006}. Next, we identify a sufficiently simple Hamiltonian $H_0$ whose ground state can be readily prepared and which can be adiabatically deformed into $H_q(\beta)$, thereby producing the Gibbs state. We emphasize that the adiabatic evolution is not restricted to the one-parameter family of Hamiltonians defined by $H_q(\beta)$ for arbitrary $\beta$~\cite{Somma2007}. In fact, asymptotic speedup over the classical Markov chain is only available along more general paths.

In two of the examples presented below, the speedup originates from ballistic propagation of domain walls enabled by quantum coherent motion as opposed to diffusive motion arising from classical thermal fluctuations. Since the width of the region explored by diffusion is proportional to the square root of time, we generically expect a quadratic speedup. Additional speedup is possible if diffusion in the Markov chain is suppressed, e.g., by a thermal barrier. An alternative speedup mechanism associated with quantum tunneling is uncovered in the problem of sampling from weighted independent sets on star graphs. We note that we only compare the quantum algorithm to the classical Markov chain from which it is constructed even though faster classical algorithms exist for the examples below~\cite{Ferris2012}. Ultimately, we are interested in the potential of our approach to general sampling problems, where Markov chain Monte Carlo is one of few algorithms with guaranteed convergence.

Our construction of the parent Hamiltonian follows the prescription in~\cite{Verstraete2006} (see~\cite{Aharonov2003,Henley2004,Castelnovo2005} for related earlier work). We first define a Markov chain that samples from the desired Gibbs distribution $p(s)$. The Markov chain is specified by a generator matrix $M$, where the probability distribution $q_t(s)$ at time $t$ evolves according to $q_{t+1}(s) = \sum_{s'} q_t(s') M(s', s)$. We assume in addition that the Markov chain satisfies detailed balance, which can be expressed as $e^{- \beta H_c(s')} M(s', s) = e^{- \beta H_c(s)} M(s, s')$.  This property implies that $p(s)$ is a stationary distribution of the Markov chain and therefore constitutes a left eigenvector of $M$ with eigenvalue unity. Moreover,
\begin{equation}
  H_q(\beta) = n \left( \mathbb{I} - e^{-\beta H_c / 2} M e^{\beta H_c/2} \right)
  \label{eq:parent}
\end{equation}
is a real, symmetric matrix and thus a valid quantum Hamiltonian. The spectrum of $H_q(\beta)$ is bounded from below by $0$ since the spectrum of the stochastic matrix $M$ is bounded from above by $1$. Furthermore, $H_q(\beta) \ket{\psi(\beta)} = 0$ such that the Gibbs state is a ground state. The factor of $n$ in \eqref{eq:parent} ensures that the spectrum of the parent Hamiltonian is extensive. To account for the natural parallelization in adiabatic evolution, we divide the mixing time of the Markov chain by $n$ for a fair comparison, denoting the result by $t_m$. The correspondence between the spectra of $M$ and $H_q(\beta)$ establishes the bound $t_m \geq 1/\Delta(\beta) - 1/n$, where $\Delta(\beta)$ is the energy gap of the parent Hamiltonian~\cite{Aldous1982,Levin2009}.

\begin{figure}[t]
  \centering
  \includegraphics{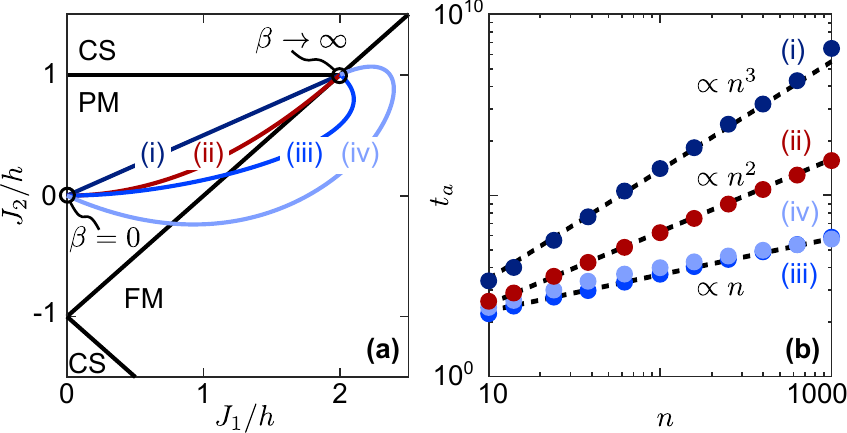}
  \caption{(a)~Phase diagram of the parent Hamiltonian corresponding to the Ising chain. The black lines indicate the boundaries between paramagnetic (PM), ferromagnetic (FM), and cluster-state-like (CS) phases. The curves labeled (i)--(iv) show four different choices of adiabatic paths with (ii) representing the one-parameter family $H_q(\beta)$. (b)~The time $t_a$ required to reach a fidelity exceeding $1-10^{-3}$ as a function of the number of spins $n$. The dashed lines are guides to the eye showing the expected linear, quadratic, and cubic relations.}
  \label{fig:fig2}
\end{figure}

We now illustrate this procedure by considering a ferromagnetic Ising model composed of $n$ spins in one dimension. The classical Hamiltonian is given by $H_c = -\sum_{i=1}^n \sigma_i^z \sigma_{i+1}^z$ with periodic boundary conditions. Choosing Glauber dynamics as the Markov chain~\cite{Glauber1963}, the corresponding parent Hamiltonian takes the form
\begin{align}
  H_q(\beta) = \frac{n}{2} \mathbb{I} -\sum_{i=1}^n \left[ h(\beta) \sigma_i^x \right. &+ J_1(\beta) \sigma_i^z \sigma_{i+1}^z \nonumber\\
  &- \left. J_2(\beta) \sigma_{i-1}^z \sigma_i^x \sigma_{i+1}^z \right],
  \label{eq:ising}
\end{align}
    where $4 h(\beta) = 1+1/\cosh(2 \beta)$, $2 J_1(\beta) = \tanh(2 \beta)$, and $4 J_2(\beta) = 1 - 1/\cosh(2 \beta)$ (see~\cite{PRA} for details and~\cite{Felderhof1971,Siggia1977} for early derivations of this result). At infinite temperature ($\beta = 0$), we have $J_1 = J_2 = 0$ and $h = 1/2$. The ground state is a paramagnet aligned along the $x$-direction, which corresponds to an equal superposition of all classical spin configurations, consistent with the Gibbs distribution at infinite temperature. When the temperature is lowered, the parameters move along a segment of a parabola in the two-dimensional parameter space $(J_1/h, J_2/h)$ shown by the red curve (ii) in \figref{fig:fig2}(a).

The quantum phase diagram of the parent Hamiltonian for arbitrary values of $h$, $J_1$, and $J_2$ is obtained by performing a Jordan--Wigner transformation that maps \eqref{eq:ising} onto a free-fermion model~\cite{Skrovseth2009,Niu2012,PRA}. The distinct quantum phases are displayed in \figref{fig:fig2}(a). The model reduces to the transverse field Ising model on the $J_2/h = 0$ axis, in which a phase transition from a paramagnet to a ferromagnet occurs at $J_1/h = 1$~\cite{Pfeuty1970}. Along the $J_1/h = 0$ axis, the ground state undergoes a symmetry-protected topological phase transition at $J_2/h = \pm 1$ from the paramagnet to a cluster-state-like phase~\cite{Son2011,Verresen2017}. We note that the tricritical point at $(J_1/h, J_2/h) = (2,1)$ describes the parent Hamiltonian corresponding to the zero temperature Gibbs distribution.

To prepare the Gibbs state $\ket{\psi(\beta)}$, one may start from the ground state of $H_q(0)$ before smoothly varying the parameters $(h, J_1, J_2)$ to bring the Hamiltonian into its final form at the desired inverse temperature $\beta$. States with finite $\beta$ can be connected to the infinite temperature state by a path that lies fully in the paramagnetic phase. Both adiabatic state preparation and the Markov chain are efficient in this case. Indeed, it has been shown previously that there exists a general quantum algorithm with run time $\sim \log n$ for gapped parent Hamiltonians~\cite{Ge2016}, which is identical to the Markov mixing time $t_m$ for the Ising chain~\cite{Levin2009}.

Sampling at zero temperature is more challenging with the mixing time of the Markov chain bounded by $t_m \gtrsim n^2$~\cite{PRA}. For the quantum algorithm, we consider the four different paths in \figref{fig:fig2}(a). To evaluate the dynamics quantitatively, we choose the rate of change of the Hamiltonian parameters with the aim of satisfying the adiabatic condition at every point along the path and numerically integrate the Schrödinger equation with the initial state $\ket{\psi(0)}$ to obtain $\ket{\phi(t_\mathrm{tot})}$ after total evolution time $t_\mathrm{tot}$ (see~\cite{PRA} for details on the adiabatic schedule). We emphasize that the adiabatic schedule proceeds with a non-constant rate, leading to scaling of the adiabatic state preparation time distinct form the Landau--Zener result~\cite{Zener1932}. To determine the dependence on the number of spins, we extract the time $t_a$ at which the fidelity $\mathcal{F} = | \inner{\phi(t_\mathrm{tot})}{\psi(\infty)}|^2$ exceeds $1-10^{-3}$ [\figref{fig:fig2}(b)]. We find three different scalings of the time $t_a$: along path (i), it roughly scales as $t_a \sim n^3$, along (ii) as $t_a \sim n^2$, while (iii) and (iv) exhibit a scaling close to $t_a \sim n$.

These scalings follow from the nature of the phase transitions. The dynamical critical exponent at the tricritical point is $z=2$, meaning that the gap closes with system size as $\Delta \sim 1/n^2$, which is consistent with the time required along path (ii). The dynamical critical exponent at all phase transitions away from the tricritical point is $z=1$ and the gap closes as $\Delta \sim 1/n$~\cite{PRA}. Therefore, the paramagnetic to ferromagnetic phase transition can be crossed adiabatically in a time proportional to $n$, only limited by ballistic propagation of domain walls as opposed to diffusive propagation in the Markov chain.  There is no quadratic slowdown as paths (iii) and (iv) approach the tricritical point, which we attribute to the large overlap of the final state with ground states in the ferromagnetic phase. Path (i) performs worse than path (ii) because the gap between the paramagnetic and cluster-state-like phases vanishes exactly for certain parameters even in a finite-sized system~\cite{PRA}. To support the claim that the speedup is quantum mechanical, we note that the half-chain entanglement entropy of the ground state diverges logarithmically with $n$ when paths (iii) and (iv) cross from the paramagnetic into the ferromagnetic phase. It is impossible to represent this ground state as a Gibbs state of a local, classical Hamiltonian $H_c$ because any such representation would be a matrix product state with constant bond dimension and bounded entanglement entropy~\cite{Verstraete2006}.

While the previous example illustrates a mechanism for quantum speedup, sampling from large systems is hard only at zero temperature~\cite{Dyer2004}, where more suitable optimization algorithms may exist. In addition, the parent Hamiltonian, \eqref{eq:ising}, does not have a simple physical realization. We address these limitations by considering the weighted independent set problem. An independent set of a graph is any subset of vertices in which no two vertices share an edge. We say vertex $i$ is occupied ($n_i = 1$) if it is in the independent set and unoccupied ($n_i = 0$) otherwise. In the maximum weighted independent set problem, each vertex is further assigned a weight $w_i$ and we seek to  minimize the energy $H_c = - \sum_i w_i n_i$ subject to the independent set constraint. The corresponding Gibbs distribution has been studied extensively~\cite{Sly2010,Gaunt1965,Weigt2001}. To construct a quantum algorithm that samples from this Gibbs distribution, each vertex is associated with a spin variable $\sigma_i^z = 2 n_i - 1$. Single spin flips with the Metropolis--Hastings update rule~\cite{Hastings1970} yield the parent Hamiltonian
\begin{align}
  H_q(\beta) = \sum_i P_i &\left[ V_{e,i}(\beta) n_i + \right. \nonumber\\
    & \phantom{[} \left. V_{g,i}(\beta) (1 - n_i) - \Omega_i(\beta) \sigma_i^x \right],
  \label{eq:mis}
\end{align}
where we only consider the subspace spanned by the independent sets. In \eqref{eq:mis}, $P_i= \prod_{j \in \mathcal{N}_i} (1 - n_j)$ projects onto states in which all nearest neighbors $\mathcal{N}_i$ of vertex $i$ are unoccupied. The parameters are given by $V_{e,i}(\beta) = e^{- \beta w_i}$, $V_{g,i}(\beta) = 1$, and $\Omega_i(\beta) = e^{- \beta w_i/2}$~\cite{PRA}.

\begin{figure}[t]
  \centering
  \includegraphics{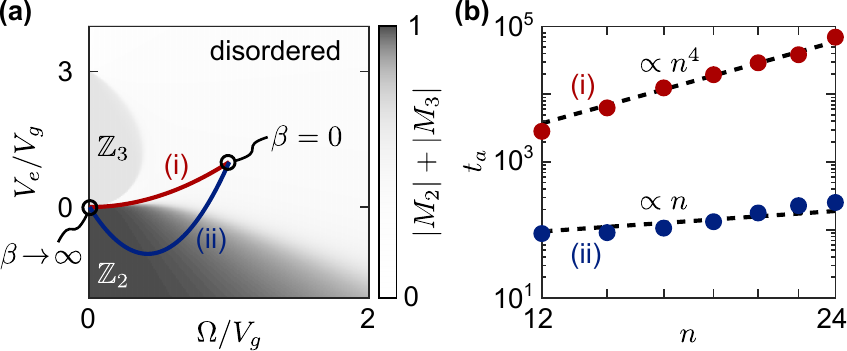}
  \caption{(a)~Parameter space and order parameter of the parent Hamiltonian for a chain of length $n=30$. The order parameter $|M_2| + |M_3|$ distinguishes the disordered phase from the $\mathbb{Z}_2$ and $\mathbb{Z}_3$ ordered phases. The red curve (i) indicates the one-parameter family $H_q(\beta)$, while the blue curve (ii) is an alternative adiabatic path crossing into the $\mathbb{Z}_2$ phase. (d)~Adiabatic state preparation time $t_a$ to reach a fidelity $\mathcal{F} > 1 - 10^{-3}$ along the two paths in (b). Path (i) terminates at $\beta_c = 2 \log n$ while the end of (ii) corresponds to the parent Hamiltonian with $\beta \to \infty$. The black lines are guides to the eye showing the scalings $t_a \propto n$ and $t_a \propto n^4$.}
  \label{fig:fig3}
\end{figure}

The projectors $P_i$ involve up to $d$-body terms, where $d$ is the degree of the graph. Nevertheless, they can be implemented with minimal experimental overhead for certain classes of graphs. In the case of so-called unit disk graphs, these operators are naturally realizable using highly excited Rydberg states of neutral atoms by extending existing schemes that implement Hamiltonians for the independent set problem using Rydberg blockade~\cite{Pichler2018a,PRA}. As a simple example of a unit disk graph, we consider a chain of length $n$ and choose equal weights $w_i = 1$. The resulting parent Hamiltonian has been studied both theoretically~\cite{Fendley2004,Lesanovsky2012} and experimentally using Rydberg atoms~\cite{Labuhn2016,Bernien2017}. Its quantum phases can be characterized by the staggered magnetization $M_k = (1/n) \sum_{j=1}^n e^{2 \pi i j / k} \sigma_j^z$. \Figref{fig:fig3}(a) shows the ground state expectation value of $|M_2| + |M_3|$ for $n=30$, clearly indicating the presence of three distinct phases. For large $\Omega/V_g$ or large, positive $V_e/V_g$, assuming $V_g > 0$ throughout, the ground state respects the full translational symmetry of the Hamiltonian and $|M_k|$ vanishes for all integers $k > 1$. When $V_e/V_g$ is sufficiently small, the ground state is $\mathbb{Z}_2$ ordered with every other site occupied and $|M_2| \neq 0$. Owing to next-to-nearest neighbor repulsive terms in the Hamiltonian, there further exists a $\mathbb{Z}_3$ ordered phase, in which $|M_3| \neq 0$.

The one-parameter family $H_q(\beta)$ is indicated by the red curve (i) in \figref{fig:fig3}(a). Although $\ket{\psi(0)}$ is not a product state, it can be efficiently prepared. For example, $H_q(0)$ can be adiabatically connected to $\Omega/V_g = 0$ and $V_e/V_g > 3$, where the ground state is the state of all sites unoccupied, by a path that lies fully in the disordered phase. Similarly, the Markov chain at infinite temperature is efficient as the parent Hamiltonian is gapped. Numerical results indicate that the gap is proportional to $e^{- 2\beta}$ at high temperature and $e^{- \beta}/n^2$ at low temperature~\cite{PRA}. The Markov chain is not ergodic at zero temperature because defects in the $\mathbb{Z}_2$ ordering, i.e., adjacent unoccupied sites, must overcome an energy barrier to propagate. It is nevertheless possible to sample approximately from the ground state by running the Markov chain at a low temperature $\beta \gtrsim \beta_c$, where $\beta_c = 2 \log n$ is the temperature at which the correlation length is comparable to the system size. The gap of the parent Hamiltonian bounds the mixing time by $t_m \gtrsim e^{2 \beta_c} = n^4$. As shown in \figref{fig:fig3}(b), the adiabatic state preparation time $t_a$ along the one-parameter family $H_q(\beta)$ follows the same scaling (see~\cite{PRA} for details concerning the adiabatic schedule).

A quantum speedup is obtained by choosing a different path. For example, \figref{fig:fig3}(b) shows an approximately linear scaling of $t_a$ along path (ii) in \figref{fig:fig3}(a). We emphasize that unlike path (i), path (ii) ends at the parent Hamiltonian $H_q(\beta)$ with $\beta \to \infty$. The quantum algorithm is thus capable of preparing the zero-temperature Gibbs state despite the lack of ergodicity of the Markov chain. We again attribute the linear scaling to the dynamical critical exponent $z = 1$ at the phase transition between the disordered and the $\mathbb{Z}_2$ ordered phases. Note that for the independent set problem, the quantum speedup is quartic owing to the more slowly mixing Markov chain. However, it is possible to improve the performance of the Markov chain by adding simultaneous spin flips on neighboring sites, reducing the advantage of the quantum algorithm to a quadratic speedup similar to the Ising model.

\begin{figure}[t]
  \centering
  \includegraphics{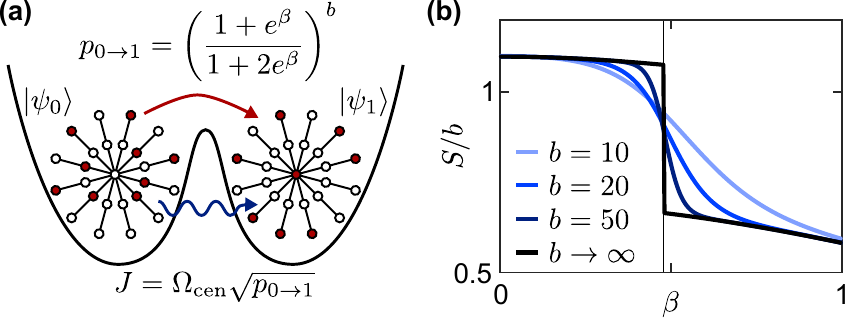}
  \caption{(a)~Sampling from a star graph with two vertices per branch. The mixing of the Markov chain is limited at low temperature by the probability $p_{0 \to 1}$ of changing the central vertex from unoccupied to occupied. The quantum algorithm achieves a quadratic speedup over the Markov chain by tunneling between such configurations with rate $J$. (b)~Entropy per branch $S/b$ of the Gibbs distribution of weighted independent set problem for this graph. The system exhibits a discontinuous phase transition at $\beta^* \approx 0.48$. The central vertex is occupied with high probability when $\beta > \beta^*$ and unoccupied otherwise.}
  \label{fig:fig4}
\end{figure}

We next consider a graph for which it is hard to sample from independent sets even at nonzero temperature. The graph takes the shape of a star with $b$ branches and two vertices per branch [\figref{fig:fig4}(a)]. The weight of the vertex at the center is $b$, while all other weights are set to 1. The classical model exhibits a phase transition at $\beta^* = \log \varphi \approx 0.48$, where $\varphi$ is the golden ratio [\figref{fig:fig4}(b)]~\cite{PRA}. The Markov chain on this graph is subject to severe kinetic constraints since changing the central vertex from unoccupied to occupied requires all neighboring vertices to be unoccupied. Assuming that each individual branch is in thermal equilibrium, the probability of accepting such a move is given by $p_{0 \to 1} = [(1 + e^{\beta})/(1 + 2 e^{\beta})]^{b}$. Similarly, the reverse process is energetically suppressed with an acceptance probability $p_{1 \to 0} = e^{-b \beta}$. The central vertex can thus become trapped in the thermodynamically unfavorable configuration, resulting in a mixing time that grows exponentially with $b$ at any finite temperature. When starting from a random independent set, the Markov chain will nevertheless sample efficiently at high temperature because the probability of the central vertex being initially occupied is exponentially small. By the same argument, the Markov chain almost certainly starts in the wrong configuration in the low temperature phase and convergence to the Gibbs distribution requires a time $t_m \gtrsim 1/p_{0 \to 1}$.

The corresponding quantum dynamics are captured by a two-state model formed by $\ket{\psi_0(\beta)}$ and $\ket{\psi_1(\beta)}$, which are Gibbs states with the central vertex fixed to be respectively unoccupied or occupied [\figref{fig:fig4}(a)]. The tunneling rate between these states, i.e.~the matrix element $\bra{\psi_0} H_q \ket{\psi_1}$, is given by $J = \Omega_\mathrm{cen} \sqrt{p_{0 \to 1}}$, where $\Omega_\mathrm{cen}$ denotes the coefficient $\Omega_i$ in \eqref{eq:mis} associated with the central vertex. The time required to adiabatically cross the phase transition is bounded by $t_a \gtrsim 1/J$ with $J$ evaluated at the phase transition. Along the one-parameter family $H_q(\beta)$, we have $\Omega_\mathrm{cen} = \sqrt{p_{1 \to 0}}$ such that adiabatic state preparation yields the same time complexity $t_a \gtrsim 1/p_{0 \to 1}$ as the Markov chain that samples at the phase transition ($p_{0 \to 1} = p_{1 \to 0}$ at the phase transition). However, the square-root dependence of the tunneling rate on $p_{0\to 1}$ suggests that a quadratic speedup may be attainable with a path along which $\Omega_\mathrm{cen} = 1$ when crossing the phase transition. An example of such a path is provided in~\cite{PRA} together with a numerical demonstration of the quadratic speedup.

Our approach to quantum sampling algorithms unveils a connection between computational complexity and phase transitions and provides physical insight into the origin of quantum speedup. The quantum Hamiltonians appearing in the construction are guaranteed to be local given that the Gibbs distribution belongs to a local, classical Hamiltonian and that the Markov chain updates are local. Consequently, time evolution under these quantum Hamiltonians can be implemented using Hamiltonian simulation~\cite{Lloyd1996}. Moreover, a hardware efficient implementation suitable for recently demonstrated two-dimensional Rydberg atom arrays~\cite{Ebadi2020,Scholl2020} is possible for certain problems~\cite{PRA}.

Further research is required to apply our approach to practically relevant problems such as disordered systems in two or more dimensions. Our work may be extended to cover quantum algorithms derived from Markov chains with cluster updates, which are often effective in practice. Moreover, it will be necessary to develop methods for efficient state preparation when computation of the full phase diagram is classically intractable, as one expects for generic instances. Hybrid algorithms, which combine quantum evolution with classical optimization, such as the recently proposed variational quantum adiabatic algorithm~{\cite{Schiffer2021}}, are particularly promising in this context. Apart from testing such algorithms, their realization on near-term quantum devices can open the door to exploration of novel applications in areas ranging from physical science to machine learning.

\begin{acknowledgments}
    We thank J.~I.~Cirac, E.~A.~Demler, E.~Farhi, A.~Polkovnikov, and P.~Zoller for insightful discussions. This work was supported by the National Science Foundation, the MIT--Harvard Center for Ultracold Atoms, the Department of Energy, and the DARPA ONISQ program. D.S. was supported by AFOSR: Grant FA9550-21-1-0236. The Flatiron Institute is a division of the Simons Foundation.
\end{acknowledgments}

\bibliography{bibliography}

\end{document}